\documentclass[reprint,aps,floatfix,footinbib,prx]{revtex4-2}
\usepackage{cancel}
\usepackage{amsmath}
\usepackage{gensymb}
\usepackage{physics}
\usepackage{graphicx}
\usepackage{amssymb}
\usepackage{amsthm}
\usepackage{bm}
\usepackage{dcolumn}
\usepackage{braket}
\usepackage{longtable}
\usepackage{tabularx}

\usepackage{ragged2e}
\usepackage{txfonts}
\usepackage{lipsum}

\usepackage{color}
\usepackage[usenames,dvipsnames]{xcolor}
\definecolor{myblue}{rgb}{0,0,1}
\usepackage[breaklinks=true,colorlinks=true,linkcolor=myblue,urlcolor=myblue,citecolor=myblue]{hyperref}

\newcommand{\vq}{{\bm{q}}}

\newcommand{\vp}{\hat{{\bm{p}}}}
\newcommand{\vk}{{\bm{k}}}
\newcommand{\vQ}{{\bm{Q}}}
\newcommand{\vkappa}{{\boldsymbol\kappa}}

\newcommand{\vm}{{\bm{m}}}
\newcommand{\vn}{{\bm{n}}}

\newcommand{\vx}{{\hat{\bm{x}}}}

\usepackage{chemformula}

\allowdisplaybreaks

\begin{document}

\title{Theory of acoustic polarons in the two-dimensional SSH model applied to the layered superatomic semiconductor \ch{Re6Se8Cl2}}

\author{Petra Shih}
\author{Timothy C. Berkelbach}
\email{tim.berkelbach@gmail.com}
\affiliation{Department of Chemistry,
Columbia University, New York, New York 10027, USA}

\begin{abstract}
Layered superatomic semiconductors, whose buildings blocks are atomically precise molecular clusters, exhibit interesting electronic and vibrational properties. 
In recent work [Science \textbf{382}, 438 (2023)], transient reflection microscopy revealed quasi-ballistic exciton dynamics in \ch{Re6Se8Cl2}, which was attributed to the formation of polarons due to coupling with acoustic phonons.
Here, we characterize the electronic, excitonic, and phononic properties with periodic density functional theory.
We further parameterize a polaron Hamiltonian with nonlocal [Su-Schrieffer-Heeger (SSH)] coupling to acoustic phonon to study the polaron ground state binding energy and dispersion relation with variational wavefunctions. 
We calculate a polaron binding energy of about 10~meV at room temperature, and the maximum group velocity of our polaron dispersion relation is 1.5~km/s, which is similar to the experimentally observed exciton transport velocity.
\end{abstract}

\maketitle
\section{Introduction}

Using precise molecular clusters to form extended materials allows the design of so-called superatomic solids with tunable crystal structures and vibrational, electronic, and magnetic properties~\cite{wang2021boosting, roy2013nanoscale, turkiewicz2014assembling, meirzadeh2023few, lee2014ferromagnetic}.
In particular, superatomic solids made from inorganic octahedral clusters have been shown to possess interesting properties such as doping-induced superconductivity~\cite{telford2020doping, fischer1973superconductivity}, an anisotropic optical response~\cite{handa2023anisotropically}, (photo)electrocatalytic activity~\cite{gaoxiang2019tuning, shoushou2023functional}, and the ability for exfoliation~\cite{zhong2018superatomic}.
Within this class of materials, the van der Waals layered semiconductor \ch{Re6Se8Cl2} has very recently been found to exhibit long-range, quasi-ballistic exciton transport after photoexcitation at room temperature~\cite{tulyagankhodjaev2023room}. 
Such transport behavior has not been reported in any other semiconductor and is rather counterintuitive because of the material's relatively flat electronic band structure.

In Ref.~\onlinecite{tulyagankhodjaev2023room}, the quasi-ballistic exciton transport, measured directly via transient reflection microscopy, was attributed to the formation of polarons.
Strong electron-phonon interactions have previously been implicated in the
temperature-dependent valence band narrowing of
\ch{Re6Se8Cl2}~\cite{li2020strong} and in the superconductivity of the related Chevrel
phase material \ch{PbMo6S8}~\cite{chen2018intermolecular}.
Moreover, the observed speed of the quasi-ballistic exciton transport was similar to that of the speed of sound of the lattice,
suggesting that the polaron were formed through interaction with acoustic phonons.
As co-authors of Ref.~\onlinecite{tulyagankhodjaev2023room}, we provided a heuristic
theory of acoustic polaron formation in \ch{Re6Se8Cl2}, and in this work, we present a more
detailed study.
We use periodic density functional theory (DFT) to characterize the electronic, excitonic, and vibrational properties of \ch{Re6Se8Cl2}, which are then combined to
study the ground-state properties of polarons.

Specifically, we study the polaron behavior of a parameterized two-dimensional (2D) Su-Schrieffer-Heeger (SSH) Hamiltonian \cite{su1979solitons} with nonlocal carrier coupling to acoustic phonons. 
Although the theory and simulation of polarons in model Hamiltonians is relatively mature~\cite{frohlich1954electrons, holstein1959studies, yarkony1977variational, venzl1985theory, peeters1985acoustical, hahn2021diagrammatic}, our Hamiltonian is especially challenging because the coupling vertex depends on both the exciton and phonon momenta, and the acoustic phonon spectrum is gapless. 
This problem of a massive carrier coupled to bosonic excitations with acoustic dispersion has been studied in the context of electrons in conjugated polymers~\cite{su1979solitons, takayama1980continuum}, vibrational excitations in DNA~\cite{davydov1973theory, davydov1979solitons}, and---more recently---quantum impurities in degenerate Bose gases~\cite{tempere2009feynman, Seetharam2021}.
In this work, we focus on the properties of the polaron ground state at zero and finite momentum, which we study with variational wavefunctions.

\section{Electronic and excitonic band structures}

\begin{figure}[t]
    \centering
    \includegraphics[scale=0.24]{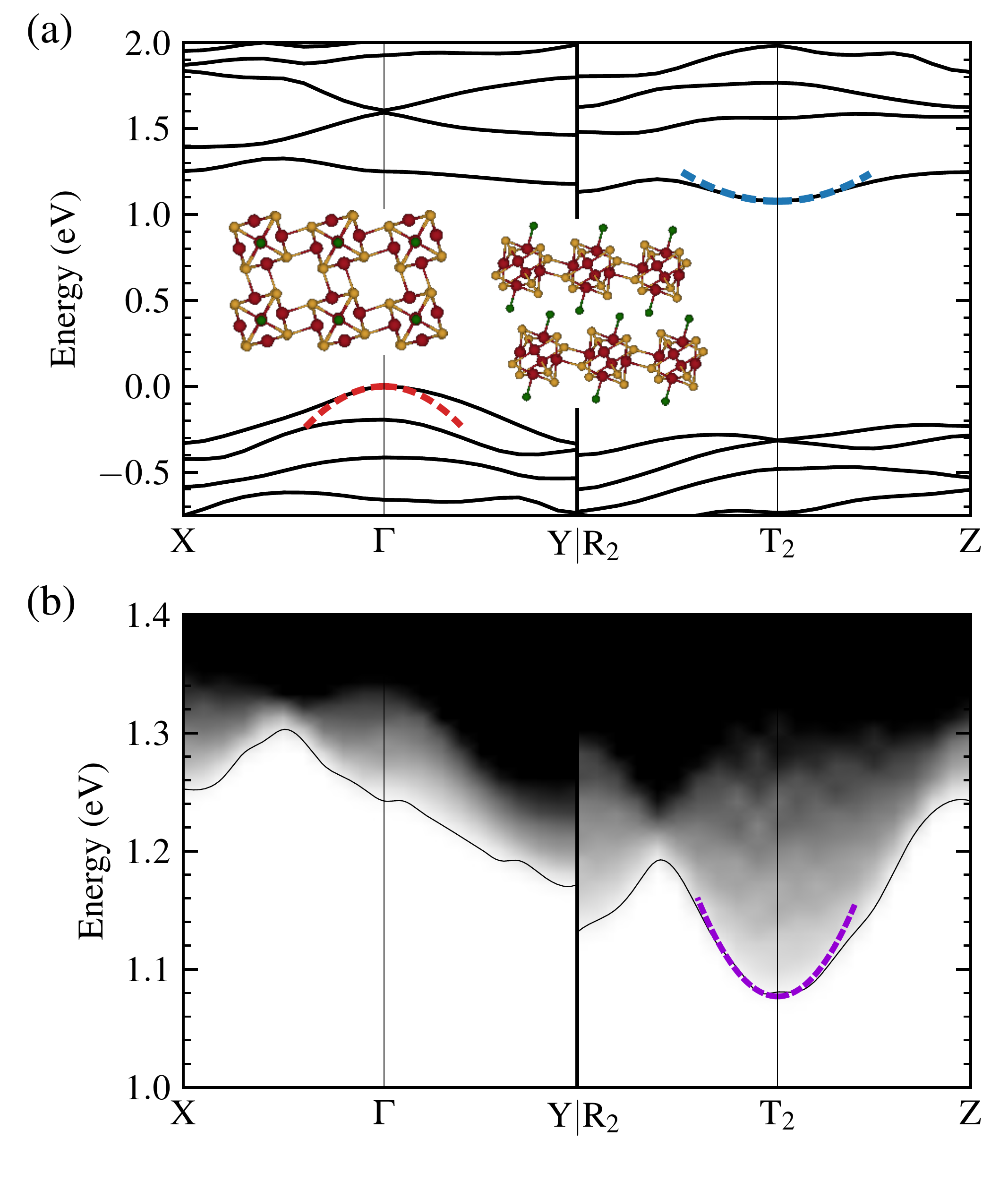}
    \caption{(a) Electronic band structure and (b) excitonic band structure, neglecting electron-hole interactions. 
The electron, hole, and exciton effective masses are $m^{\ast}_\mathrm{e} = 1.26$~$m_0$, $m^{\ast}_\mathrm{h} = 0.61$~$m_0$, and $m_\mathrm{ex}^{\ast} = m^{\ast}_\mathrm{e} + m^{\ast}_\mathrm{h} = 1.87$~$m_0$.
The inset of (a) shows the atomic structure of \ch{Re6Se8Cl2} with Re atoms in red, Se atoms in yellow, and Cl atoms in green.
}
    \label{fig:band_structure}
\end{figure}

\ch{Re6Se8Cl2} is a layered material; each \ch{Re6Se8} unit is covalently bonded to four neighboring units, forming a 2D lattice that is capped on both sides by chlorine atoms, as shown in the inset of Fig.~\ref{fig:band_structure}(a).
We start by characterizing the static electronic structure of \ch{Re6Se8Cl2} using
DFT calculations with the PBE exchange correlation functional and projector augmented wave pseudopotentials;
all calculations are performed using Quantum Espresso~\cite{giannozzi2009quantum, giannozzi2017advanced}.
We optimize the internal atomic positions while keeping the lattice parameters fixed at their experimental values~\cite{zhong2018superatomic}.

The calculated electronic band structure, presented in Fig.~\ref{fig:band_structure}(a) and Fig.~\ref{fig:band_structure_full}, shows that \ch{Re6Se8Cl2} is a semiconductor with an indirect band gap between the valence band maximum at $\Gamma$~[$\vk=(0, 0, 0)$] and the conduction band minimum at $\mathrm{T_{2}}$~[$\vk=(0, -\frac{1}{2}, \frac{1}{2})$], 
where $\vk$ is given in units of the reciprocal lattice vectors, and special $k$-point labels follow the HPKOT convention for the triclinic 
lattice system with acute angles between reciprocal vectors \cite{hinuma2017band}.
At this level of theory, the band gap is about 1.1~eV, although tunneling measurements and previous calculations with a hybrid functional 
suggest a band gap of about 1.6~eV~\cite{zhong2018superatomic}. 
We observe that \ch{Re6Se8Cl2} has a congested band structure with relatively flat bands, reflecting the weak hybridization between clusters.
We determined an in-plane effective mass for electrons and holes by fitting the band structure along the high-symmetry directions R$_2$-T$_2$-Z for the electron and X-$\Gamma$-Y for the hole. 
This yields $m^{\ast}_\mathrm{e} = 1.26$~$m_0$ and 
$m^{\ast}_\mathrm{h} = 0.61$~$m_0$, where $m_0$ is the electron mass. 
These effective masses are significantly larger than those of other two-dimensional semiconductors, such as the transition-metal dichalcogenides, which have $m^{\ast}_\mathrm{e/h} \approx 0.15$--$0.3$.

In Fig.~\ref{fig:band_structure}(b), we show the excitonic band structure in the absence of electron-hole interactions, visualized as the spectral function
\begin{equation}
A(\vQ,E) = N_k^{-1} \sum_{cv\vk} \delta(\varepsilon_c(\vk+\vQ)-\varepsilon_v(\vk)-E),
\end{equation}
where $\vQ$ is the exciton center-of-mass momentum and $c,v$ are conduction and valence
band indices. 
Consistent with its indirect bandgap, 
the lowest-energy exciton dispersion has a minimum at $\mathrm{T_{2}}$ with an
effective mass of $m_\mathrm{ex}^{\ast} = 1.87$~$m_0$, which is consistent with
the electronic band structure and effective electron and hole masses.  
However, we emphasize that excitonic effects are expected to be relatively large for this
material, due to its quasi-2D character and flat bands: using the reduced
exciton mass of $\mu_\mathrm{ex} = 0.41$~$m_0$ and the dielectric constant of
$\epsilon = 10.5$, a 2D hydrogen model of Wannier excitons predicts a binding
energy of $E_\mathrm{ex} \approx 200$~meV and radius of $a_\mathrm{ex} \approx
6$~\AA, which is about the size of a single cluster.  Together, these
calculations support a Frenkel exciton picture of strongly bound intracluster
excitons.

For the coupling to phonons, which we consider in the following sections, we use a simplified model of the excitonic band structure, which is a one-band model on a two-dimensional square lattice with nearest-neighbor 
transfer integrals of opposite sign, $J_\mathrm{x} = -J_\mathrm{y} \equiv J_0$.
This simplification is motivated by the strong 
excitonic effects, the weak interlayer interactions, and the fact that the in-plane lattice parameters 
have an angle of about 90$\degree$.
However, we emphasize that there are many nearby bands whose influence should be considered in future work.
The model exciton band energy is thus
\begin{equation}
\label{eq:exc_bands}
E^{\text{ex}}(\vQ) = -2J_0\left[ \cos(Q_\mathrm{x}a) - \cos(Q_\mathrm{y}a) \right],
\end{equation}
where $a$ is the in-plane lattice constant and $J_0 = \hbar^2/(2m_\mathrm{ex}^{\ast}a^2) = 46$~meV as determined by fitting to the excitonic band structure.

\section{Exciton-phonon interactions}

\begin{figure}[b]
    \centering
    \includegraphics[scale=0.85]{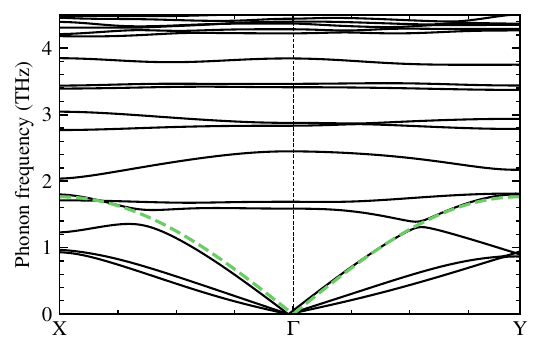}
    \caption{Phonon band structure along the path X-$\Gamma$-Y and the fitting of the the longitudinal acoustic phonon dispersion along $\Gamma$-X and $\Gamma$-Y with $\Omega_\vq = 2\Omega|\sin(qa/2)|$. An in-plane inter-cluster vibrational frequency of $\Omega \approx 0.9$~THz (3.7~meV) is obtained.}
    \label{fig:ph_band_structure_XGY}
\end{figure}
\ch{Re6Se8Cl2} has many phonon modes due to the large number of atoms in the unit cell.
In Fig.~\ref{fig:ph_band_structure_XGY}, we show a portion of the DFT phonon band structure at frequencies below 
4.5~THz (19~meV) [in Fig.~\ref{Appendix:ph_band_structure}, we show the entire phonon band structure, which exhibits phonon bands between
0~THz and 10~THz (about 0 to 40~meV)].
Eigenvector analysis reveals that the three lowest frequency optical phonons are cluster twisting modes,
which are followed at higher frequency by all of the intracluster normal modes.
We separate the phonon
modes into optical and acoustic modes and treat their coupling to excitons differently, as described in the next subsections. 

\subsection{Exciton-optical phonon coupling}

We first consider the coupling between excitons and optical phonons. 
Experimental angle-resolved photoemission spectra of \ch{Re6Se8Cl2} reported by Roy, Zhu, and co-workers~\cite{li2020strong} 
reported an unusually strong temperature dependence in the valence bandwidth $W$. 
Their data were accurately reproduced by assuming strong 
Holstein-type coupling between electrons and optical phonons,
which yields the temperature-dependent band narrowing~\cite{hannewald2004theory}
\begin{equation}
W(T)={W}_0 \exp\left\{-g^2 \left[n(\omega,T)+1/2\right]\right\},
\end{equation}
where $n(\omega,T) = \left[\exp \left(\hbar \omega / {k}_{\mathrm{B}} {T}\right)-1\right]^{-1}$ is the phonon occupancy for a mode with frequency
$\omega$ and $g$ is the effective coupling strength. 
The data were best fit with the parameters $W_0 = 280$~meV, $\omega = 2.6$~THz ($\hbar\omega = 11$~meV), and $g=1.2$~\cite{li2020strong}.
Considering our phonon band structure in Fig.~\ref{fig:ph_band_structure_XGY}, the value of $\omega$ suggests that 
the coupling is to the cluster twisting modes.

\begin{figure}[t]
    \centering
    \includegraphics[scale=0.85]{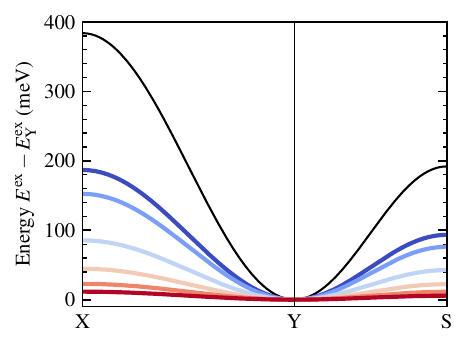}
    \caption{Lowest exciton band for 2D square lattice with temperature renormalization from low (10 K, in blue) to high (300 K, in red) temperature. $J(0$ K$) = 46$~meV, $\hbar\omega=10.8$~meV, and $g=1.2$ are used \cite{li2020strong}. }
    \label{fig:renormalization}
\end{figure}

Because we are concerned with excitons in this work, we assume the same renormalization for the conduction band.
This implies a temperature dependent exciton transfer integral $J(T)$, which yields
the temperature dependent excitonic band structure shown 
in Fig.~\ref{fig:renormalization} from 10 K to 300 K [we plot the model exciton band structure Eq.~(\ref{eq:exc_bands}) along the path X$(\frac{1}{2}, 0)$-Y$(0,\frac{1}{2})$-S$(\frac{1}{2},\frac{1}{2})$,
which corresponds roughly to the path U$_{2}(-\frac{1}{2}, 0,\frac{1}{2})$-T$_{2}(0,-\frac{1}{2},\frac{1}{2})$-R$_{2}(-\frac{1}{2},-\frac{1}{2},\frac{1}{2})$ for the triclinic lattice].
At 300 K, a transfer integral of $J(300~\mathrm{K}) = 1.4$~meV is predicted,
which is significantly smaller than 
the DFT value of $J_0 = 46$~meV 
or the 0~K value of $J(0~\mathrm{K}) = e^{-g^2/2}J_0 = 22$~meV
(the latter is reduced by zero-point vibrations).
Such a small exciton kinetic energy favors strong binding of an acoustic polaron, which might be important to understanding
the experimental transport results in Ref.~\onlinecite{tulyagankhodjaev2023room}.

Assuming an anti-adiabatic treatment of this coupling yields a temperature-dependent Hamiltonian, with a transfer
integral $J(T)$,
which we use for the rest of this work.
However, we note that the theoretical justification for an anti-adiabatic treatment leading to bandwidth narrowing is questionable, considering the ratio $J_0/\hbar\omega \approx 4$.
We therefore suggest $J = 1.4$~meV to be a lower bound at room temperature; 
the ``correct'' value of $J$ may be larger and the optical phonons arising from cluster twisting may need
to be treated dynamically.

\subsection{Exciton-acoustic phonon coupling}

The coupling to low-frequency acoustic phonons surely cannot be eliminated in the same way,
and in this work we model the coupling explicitly and use variational wavefunctions to evaluate
properties.
Specifically, we consider an exciton on a 2D lattice coupled to two branches of acoustic phonons
as described by the SSH Hamiltonian,
\begin{equation}\label{eq:SSH-Hamiltonian}
\begin{split}
H &= -\sum_{\vn,\vm}^{\prime}\left[J_{\vn\vm}-\alpha\left(\vx_\vm-\vx_\vn\right)\right]\left(a_\vn^{\dagger} a_{\vm}+\text { H.c. }\right)\\
&\hspace{1em}+\sum_\vn\left[\frac{\vp_{\vn}^2}{2 M}+\sum_{\vm}^{\prime}\frac{1}{2} M \Omega^2\left(\vx_{\vm}-\vx_\vn\right)^2\right] \\
&=\sum_\vQ E^{\text{ex}}(\vQ) a_\vQ^{\dagger} a_\vQ
    + \sum_{\nu,\vQ,\vq} g^{\nu}({\vQ,\vq}) a_{\vQ+\vq}^{\dagger} a_{\vQ} 
        (b^{\nu}_{\vq}+b_{-\vq}^{\nu\dagger}) \\
    &\hspace{1em} + \sum_{\nu,\vq}\hbar\Omega^{\nu}_{\vq}b_{\vq}^{\nu\dagger}b^{\nu}_{\vq},
\end{split}
\end{equation}
where $\vx_{\vn}$ and $\vp_{\vn}$ are the 2D lattice displacement and momentum operators at site $\vn$, 
$a^{\dagger}_{\vQ}$ is an exciton creation operator, 
$b^{\nu\dagger}_{\vq}$ is a phonon creation operators for mode $\nu=x,y$,
and the primed summation indicates a restriction to nearest neighbors on the lattice.
The SSH exciton-phonon coupling and acoustic phonon dispersion are
\begin{align}
g^{\nu}({\vQ,\vq}) &= -\frac{2i\alpha}{N^{1/2}}\left\{\sin(Q_{\nu} a) - \sin[(Q_{\nu}+q_{\nu})a] \right\} \\
\Omega^{\nu}_\vq &= 2\Omega|\sin(q_{\nu}a/2)|.
\end{align}

The cluster mass is $M=1820$~amu, and we fit the longitudinal acoustic phonon branch,
as shown in Fig.~\ref{fig:ph_band_structure_XGY}, to extract an inter-cluster vibrational
frequency of $\Omega = 0.88$~THz ($\hbar\Omega = 3.65$~meV).
Neglecting excitonic effects, we use DFT to calculate the deformation potential $D$, from which the 
exciton-phonon coupling constant is calculated as $\alpha = D/2a \approx 336$~meV/\AA.

Our use of an SSH Hamiltonian is motivated by the extensive work on molecular
solids~\cite{Troisi2006,Coropceanu2007,Wang2011,fetherolf2020unification}, in
which electrons couple to high-frequency intramolecular Holstein modes and to
low-frequency intermolecular Peierls modes. This is clearly similar to the
picture proposed in this work, and consistent with the molecular nature of the clusters
composing superatomic solids.

\section{Polaron properties}

Given the parameterized Hamiltonian in Eq.~(\ref{eq:SSH-Hamiltonian}), we seek the properties of the polaron ground state.
In Ref.~\onlinecite{tulyagankhodjaev2023room}, we offered a schematic, adiabatic strong-coupling theory of the polaron.
In that limit, the polaron is characterized exclusively by the dimensionless parameter
$\lambda = \alpha^2/JM\Omega^2$, and the polaron is predicted to be bound only if $\lambda > \pi/2$.
The smallness of $J$ due its purported temperature dependence yields a value of $\lambda \approx 14$, indicating a strongly bound acoustic polaron in support of the experimental observations.

\subsection{Variational wavefunctions}

Here, we improve upon our previous theoretical description~\cite{tulyagankhodjaev2023room} by presenting fully quantum calculations of the polaron ground state and its dispersion relation, and we explore a range of parameters to account for uncertainties in our parameterization.
From a numerical point of view, our Hamiltonian is one of the hardest polaron problems to study
due to the nonlocal form of the interaction $g^\nu(\vQ,\vq)$ and the gapless nature of the
acoustic phonon dispersion.
To overcome these challenges, in this work, we use two variational
wavefunctions, whose energies are minimized numerically. 
All calculations are performed on a periodic 2D lattice. Convergence with respect to lattice size has been tested,
and our results are presented for a 20$\times$20 lattice unless otherwise specified.

The first variational wavefunction is a localized, strong coupling ansatz,
in the spirit of the Pekar or Davydov ansatzes,
\begin{equation}
\begin{aligned}
|\Psi_{\mathrm{loc}}\rangle
=\sum_\vn \psi_\vn a_\vn^{\dagger}|0\rangle_{\mathrm{ex}} 
&\otimes\prod_{\vn} e^{-i\bm{x}_\vn\cdot\vp_{\vn}}|0\rangle_{\mathrm{ph}},
\end{aligned}
\end{equation}
where $\psi_\vn$ and $\bm{x}_{\vn}$ are variational parameters corresponding to the exciton wavefunction 
amplitudes and lattice displacements; the latter equivalence can be seen from the
coherent state property $\langle \Psi_\mathrm{loc} | \vx_\vn | \Psi_\mathrm{loc}\rangle = \bm{x}_\vn$.
We variationally minimize the energy
\begin{equation}
\label{eq:eloc}
E_\mathrm{loc} = \langle \Psi_\mathrm{loc} | H |\Psi_\mathrm{loc}\rangle 
    = \langle \bar{H} \rangle_\mathrm{ex} + \sum_{\vn,\vm}^{\prime}\frac{1}{2} M \Omega^2\left(\bm{x}_{\vm}-\bm{x}_\vn\right)^2,
\end{equation}
where $\langle \bar{H} \rangle_\mathrm{ex}$ is the ground-state eigenvalue of the purely excitonic Hamiltonian
\begin{equation}
\bar{H} = -\sum_{\vn,\vm}^{\prime}\left[J_{\vn\vm}-\alpha\left(\bm{x}_{\vm}-\bm{x}_\vn\right)\right]\left(a_\vn^{\dagger} a_{\vm}+\mathrm{H.c.}\right).
\end{equation}
This localized wavefunction is an adiabatic version of the more general Davydov ansatz~\cite{davydov1977solitons,scott1992davydov}, as can be seen from the
absence of the phonon kinetic energy in Eq.~(\ref{eq:eloc}).
However, numerical testing (not shown) indicates that the addition of phonon momenta as variational parameters does not significantly alter our results.

The localized ansatz is not generally an eigenstate of the lattice translation operator, which has a number of
unfortunate, related implications.
First, the polaron dispersion relation cannot be a straightforwardly obtained, although approximate formulations exist~\cite{whitfield1976interaction}.
Second, the optimized wavefunctions break translation symmetry, and the exciton wavefunction and lattice displacement colocalize 
around an arbitrary lattice site.
Finally, the polaron is only bound above a critical effective coupling strength (numerically, we find this
critical coupling strength to be identical to the one we derived schematically in 
Ref.~\onlinecite{tulyagankhodjaev2023room}, i.e., $\lambda > \pi/2$).
These latter behaviors, sometimes referred to as self-trapping, are likely artifacts of 
the variational wavefunction.
The absence of polaron self-trapping has been proven mathematically for a variety of 
electron-phonon Hamiltonians with Fr\"olich-like coupling to optical phonons~\cite{Spohn1987,gerlach1991analytical};
however, to the best of our knowledge, no such universal proofs exists for problems with acoustic phonons or
nonlocal coupling with a vertex that depends on both the electron and phonon momentum,
both of which apply in our model. 

A better variational wavefunction without these deficiencies is 
a delocalized one known as
Toyozawa's ansatz \cite{toyozawa1961self, zhao1997variational},
which can be obtained as a translational-symmetry adapted superposition of localized wavefunctions, leading to
\begin{equation}
\begin{aligned}
&|\Psi_{\mathrm{deloc}}(\vkappa)\rangle = \braket{\vkappa|\vkappa}^{-1/2}\left|{\vkappa}\right\rangle,\\
&\left|{\vkappa}\right\rangle = N^{-1} \sum_{\vn\vQ} e^{i(\vkappa-\vQ) \cdot \vn} \psi_{\vQ}^{\vkappa} a_{\vQ}^{\dagger} |0\rangle_{\mathrm{ex}}\\
&\hspace{1em} \otimes \prod_{\nu,\vq} \exp \left[-\frac{1}{\sqrt{N}} \left(\beta^{\nu\vkappa}_{\vq} e^{-i \vq \cdot \vn} {b^\nu_{\vq}}^{\dagger}-{\beta^{\nu\vkappa}_{\vq}}^{*} e^{i \vq \cdot \vn} b^\nu_{\vq}\right)\right] |0\rangle_{\mathrm{ph}}.
\end{aligned}
\end{equation}
Here, $\vkappa$ is the total lattice momentum, $\psi_{\vQ}^{\vkappa}$ is the exciton wavefunction amplitude, 
and $\beta^{\nu\vkappa}_{\vq}$, which is complex, is the generalized phonon deformation. 
We calculate the polaron dispersion by minimizing, at each $\vkappa$, 
\begin{equation}
E_\mathrm{deloc}(\vkappa) = \braket{\Psi_{\mathrm{deloc}}(\vkappa)|H|\Psi_{\mathrm{deloc}}(\vkappa)}
\end{equation}
with respect to $\psi^{\vkappa}_{\vQ}$ and $\beta^{\nu\vkappa}_{\vq}$ subject to a constraint on the normalization of the wavefunction.
We note that the delocalized Toyozawa ansatz has $6N$ real variational parameters whereas the localized ansatz only has $4N$ real variational parameters, but as mentioned above this difference is not responsible for its improved performance.

\begin{figure}[b]
    \centering
    \includegraphics[scale=0.7]{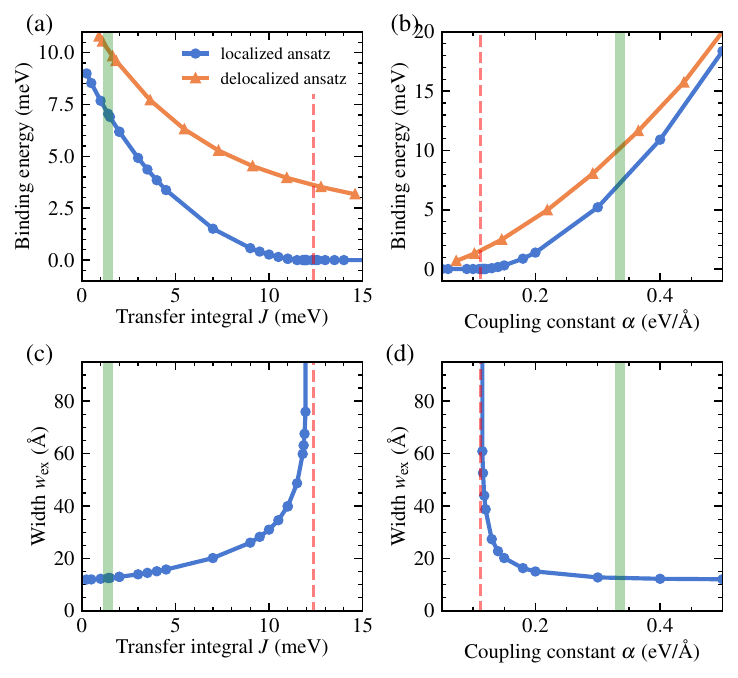}
    \caption{Polaron binding energy and width against (a,c) $J$ and (b,d) $\alpha$. A large simulation size, 80$\times$80 lattice, was used in the localized calculations to achieve critical effective coupling strength convergence. Red dashed lines denote $\lambda = \pi/2$.
}
    \label{fig:binding_energy}
\end{figure}

In Fig.~\ref{fig:binding_energy}, we show the polaron binding energy (at $\vkappa=(0,\frac{1}{2})$ for the delocalized ansatz) as
a function of the transfer integral magnitude $J$ (a) or the exciton-phonon coupling constant $\alpha$ (b),
which are varied in a range near our estimate for \ch{Re6Se8Cl2}.
As discussed, the localized ansatz binding energy goes to zero at critical values of $J$ or $\alpha$, while
the binding energy of the delocalized ansatz is always nonzero and fully analytic.
In addition to its better symmetry properties, the delocalized ansatz is a better wavefunction in the variational sense,
always predicting a binding energy that is larger in magnitude than that of the localized ansatz.
The absolute magnitude of the polaron binding energies is less than about 10~meV, which is not very large, especially considering that the quasi-ballistic transport behavior was observed at room temperature, for which $k_\mathrm{B}T \approx 26$~meV.

\begin{figure}[b]
    \centering
    \includegraphics[scale=0.7]{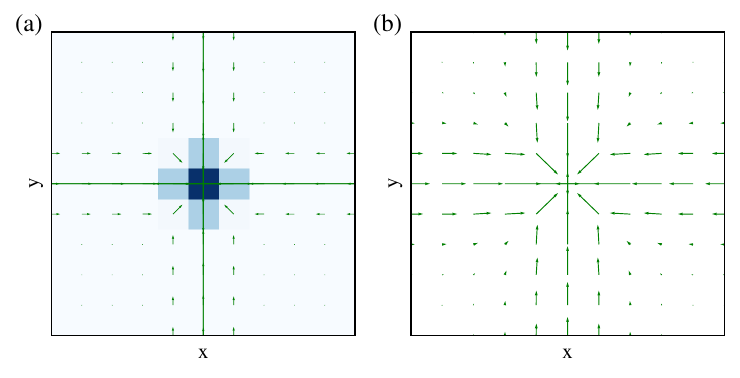}
    \caption{(a) Exciton density (blue) and phonon displacement (green) from the localized solution; (b) Exciton-phonon correlation $\bm{C}_{\vm}$ from the delocalized solution.
}
    \label{fig:polaron_correlation}
\end{figure}

In Fig.~\ref{fig:binding_energy}(c) and (d), we show the width of the exciton wavefunction in the localized ansatz~\cite{miyasaka2001acoustic},
\begin{equation}
    w_{\mathrm{ex}}=\left({\sum_{\vn} |\psi_\vn|^{4}}\right)^{-1/2}.
\end{equation}
Over most values considered here, the width is less than 20~\AA, which is only about 2--3 unit cells,
and therefore the polaron is relatively ``small''.
An analogous width cannot be calculated for the delocalized ansatz.
Instead, we can quantify the correlation between the exciton density and the lattice displacement through the correlation function

\begin{equation}
\label{eq:corr_fn}
\bm{C}_{\vm} = \sum_{\vn}\braket{a_{\vn}^{\dagger}a_{\vn}\vx_{\vn+\vm}},
\end{equation}
which we note is a vector valued quantity.
We compare the polaron structures from the localized and delocalized solutions in Fig.~\ref{fig:polaron_correlation}:
in (a), we show the exciton wavefunction and lattice displacements from the localized solution, and in (b) we show the 
correlation function~(\ref{eq:corr_fn}) from the delocalized solution. There is good qualitative agreement.
However, we find that the correlation function decays linearly with $|\vm|$, i.e., $C_\vm \sim a - b|\vm|$, such that no lengthscale can be extracted.

\subsection{Polaron dispersion relation}

\begin{figure}[b]
    \centering
    \includegraphics[scale=0.7]{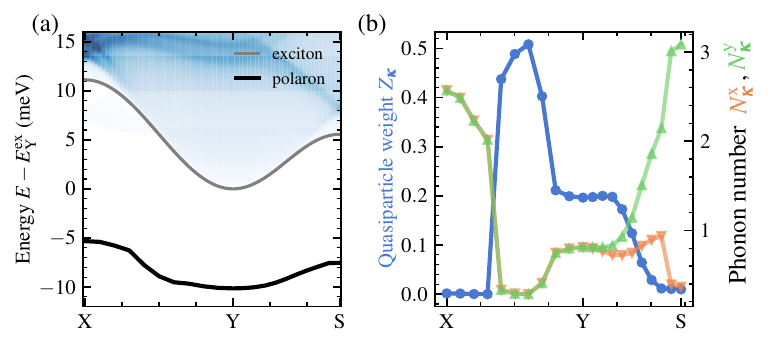}
    \caption{(a) Bare exciton dispersion (grey), polaron dispersion (black curve) and non-interacting band (blue) along the X-Y-S path, all shifted with exciton energy at Y, $E^{\text{ex}}_{\text{Y}}$. Broadening for delta function $\eta = 0.2$~meV. (b) Quasiparticle weight and phonon number along the same path.
}
    \label{fig:dispersion}
\end{figure}

We study the polaron ground-state dispersion $E(\vkappa)$ using the delocalized ansatz with general $\vkappa$. Starting from $|\Psi_{\mathrm{deloc}}(\vkappa)\rangle$, we use it as initial guess to solve for the wavefunction at neighboring $\vkappa$ points sequentially.
In Fig.~\ref{fig:dispersion}(a), we show the polaron dispersion along with the bare exciton dispersion and the continuum of a noninteracting exciton and phonons,
\begin{equation}
I(\vkappa, E) = \sum_{\vq}\sum_{m,n\ge 0} \delta(E-\varepsilon_{\vkappa-\vq}-m\Omega^{x}_{q_{x}}-n\Omega^{y}_{q_{y}}).
\end{equation}
Results are shown for the experimentally relevant parameters along the same path defined in Fig.~\ref{fig:renormalization}. 
The polaron ground-state dispersion exhibits a minimum at $\vkappa = (0, \frac{1}{2})$, i.e., the same Y point as the minimum of the exciton dispersion.

To further characterize the polaron wavefunction at all $\vkappa$, we calculate the momentum dependent exciton quasiparticle weight,
\begin{equation}
Z_{\vkappa} = \left|\bra{\Psi_{\mathrm{deloc}}(\vkappa)}a_{\vkappa}^{\dagger}\ket{0}\right|^{2}
    = \dfrac{1}{\left<\vkappa|\vkappa\right>}\left| {\psi_{\vkappa}^{\vkappa}}e^{-(2N)^{-1}\sum_{\nu, \vq} \left|\beta_\vq^{\nu \vkappa}\right|^2}\right|^{2},
\end{equation}
and phonon number
\begin{equation}
N_{\vkappa}^{\nu} = \sum_{\vq} \bra{\Psi_\mathrm{deloc}(\vkappa)} b^{\nu\dagger}_{\vq} b^{\nu}_{\vq}\ket{\Psi_\mathrm{deloc}(\vkappa)},
\end{equation}
presented in Fig.~\ref{fig:dispersion}(b).
At Y, the polaron ground state exhibits $Z_\vkappa \approx 0.2$ and roughly one phonon in each branch,
indicating strong polaron effects that deviate from those of an isolated exciton.  
Moreover, near the Brillouin zone edges, the quasiparticle weight drops abruptly to zero and the polaron 
acquires a large number of phonons. 
Achieving convergence in the delocalized variational parameters becomes more challenging in these character switching regions.
As expected, the behavior of the two phonon branches is identical along Y-X, but asymmetric along Y-S.

\begin{figure}[h]
    \centering
    \includegraphics[scale=0.7]{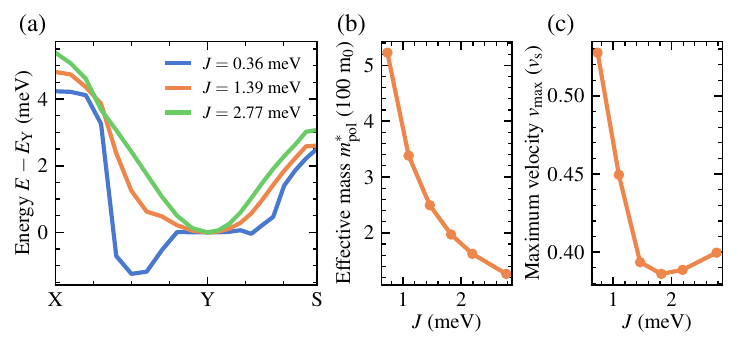}
    \caption{Polaron dispersion (a), effective mass (b), and maximum group velocity (c) for various values of the exciton transfer integral $J$. In (b), $m_0$ is the rest mass of the electron, and in (c), $v_{\mathrm{s}}$ is the speed of sound of the lattice.
}
    \label{fig:effectiveMass}
\end{figure}

Finally, we study the properties of the polaron dispersion as a function of the transfer integral $J$.
In Fig.~\ref{fig:effectiveMass}(a), we show the dispersion for three values of $J$---one smaller
and one larger than our estimated value at 300~K.
We see that with sufficiently small $J$, the minimum in the dispersion relation is shifted to a different
value of the lattice momentum, as seen in other studies of polarons with nonlocal coupling~\cite{Zhao1997nonlocal}. 
Away from this strong-coupling transition, we fit the dispersion with a quadratic function around Y and compute the polaron effective mass $m_{\text{pol}}^{\ast}=\hbar^2(\partial^{2} E(\vkappa)/\partial \kappa_\text{x}^{2}|_{\text{Y}})^{-1}$ in the $\kappa_\text{x}$ direction. Results are shown in Fig.~\ref{fig:effectiveMass}(b) as a function of $J$.
With decreasing $J$, the phonon dressing leads to a rapid increase of the effective mass from its bare value $m^\ast_{\text{ex}}$.
Specifically, we see exponential increase, $m_{\text{pol}}^{\ast} \propto \exp(c/J^2)$.
For \ch{Re6Se8Cl2}, the polaron effective mass is very large, $m_{\text{pol}}^{\ast} \approx 260 m_0$.

With an interest in understanding the quasi-ballistic transport, we evaluate the polaron group velocity using $\partial E(\vkappa)/ \partial \kappa_\mathrm{x}$. 
We recall that the maximum group velocity of our free exciton is $2Ja/\hbar$, which is between 93~km/s and 3~km/s in the absence or presence of coupling to optical phonons at room temperature.
With coupling to acoustic phonons, the maximum group velocity of the polaron is significantly depressed from these values and becomes linked to the speed of sound of the acoustic phonons.
In Fig.~\ref{fig:effectiveMass}(c), we show the maximum group velocity of the polaron along the Y-S path as a fraction of the speed of sound $v_\mathrm{s}$ and as a function of $J$. 
For the value of $J$ that we estimate for \ch{Re6Se8Cl2}, we find a maximum group velocity $v_{\mathrm{max}}
\approx 0.4 v_{\mathrm{s}} \approx 1.5$~km/s, which is similar to that observed for polaron transport in Ref.~\onlinecite{tulyagankhodjaev2023room}.

\section{Conclusions}

We have studied the electronic and vibrational properties of \ch{Re6Se8Cl2} with first principles calculations and constructed a 2D SSH model to explore the ground state properties of acoustic polarons in the material.
Using variational wavefunctions, we have demonstrated the impact of the exciton transfer integral and the exciton-phonon coupling strength on the polaron binding energy.
Within the parameter regimes we tested, a translationally symmetric delocalized ansatz always has a lower energy than a localized one, suggesting that the 2D acoustic polaron with nonlocal, SSH-type coupling shows no discontinuous self-trapping transition.

Concerning the connection to experiments,
we find that a small exciton bandwidth, which can be explained by the temperature-dependent renormalization due to optical phonons, is essential to the stability of acoustic polarons in \ch{Re6Se8Cl2}.
Within this mechanism, the polaron binding energy decreases with decreasing temperature, which aligns with the temperature-dependent transport measurements in Ref.~\onlinecite{tulyagankhodjaev2023room}. 
Moreover, the maximum group velocity of our calculated polaron dispersion is similar to the experimentally observed quasi-ballistic transport velocity, which is a bit less than the speed of sound of the lattice.

It still remains to explain the dynamical stability of the acoustic polaron and its apparent protection from scattering. As an approximate eigenstate of our model Hamiltonian, the polaron ground state has an infinite lifetime. To address these open questions requires consideration of experimentally prepared nonstationary states, finite temperature, and/or scattering mechanisms not accounted for by the model Hamiltonian.
The family of wavefunctions used in this work can be straightforwardly used to simulate such non-equilibrium quantum dynamics using the time-dependent variational principle. Work along these lines is currently in progress.

\vspace{1em}

We thank Jack Tulyagankhodjaev, Milan Delor, and David Reichman for helpful discussions.
This material is based upon work supported by the NSF MRSEC program
through Columbia in the Center for Precision Assembled Quantum Materials (DMR-2011738).
We acknowledge computing resources from
Columbia University’s Shared Research Computing Facility project, which is
supported by NIH Research Facility Improvement Grant 1G20RR030893-01, and
associated funds from the New York State Empire State Development, Division of
Science Technology and Innovation (NYSTAR) Contract C090171, both awarded April
15, 2010.

\bibliography{polaronsnew}
\pagebreak
\clearpage
\raggedbottom
\pagebreak
\widetext
\begin{center}
\textbf{\large Supplemental Material:}
\end{center}
\setcounter{section}{0}
\setcounter{equation}{0}
\setcounter{figure}{0}
\setcounter{table}{0}
\setcounter{page}{1}
\makeatletter
\renewcommand{\theequation}{S\arabic{equation}}
\renewcommand{\thesection}{S\arabic{section}}
\renewcommand{\thefigure}{S\arabic{figure}}

\section{Electronic band structure}\label{Appendix:band_structure_standard_path}
\begin{figure}[h]
    \centering
    \includegraphics[scale=0.7]{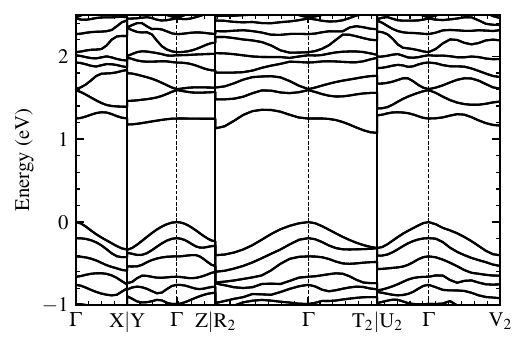}
    \caption{Electronic band structure following the HPKOT convention \cite{hinuma2017band}.}
    \label{fig:band_structure_full}
\end{figure}

\section{Phonon band structure}\label{Appendix:phonon}
\begin{figure}[h]
    \centering
    \includegraphics[scale=0.8]{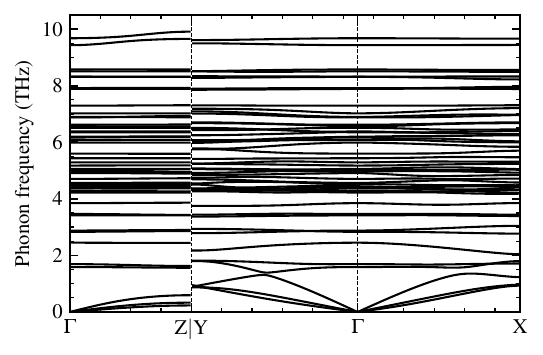}
    \caption{Phonon band structure.}
    \label{Appendix:ph_band_structure}
\end{figure}
We obtain the phonon band structure with the small displacement method implemented in the Phonopy package \cite{phonopy-phono3py-JPCM, phonopy-phono3py-JPSJ}.  A 2$\times$2$\times$2 supercell was constructed from the primitive cell parameters and atomic positions optimized with a kinetic energy cutoff of 110 Ry for wavefunction (and 360 Ry for charge density and potential) and a 4$\times$4$\times$2 $k$-point mesh to sample the first Brillouin zone using the Quantum Espresso package. The forces under small displacement are calculated at the same cutoff criteria and with a 2$\times$2$\times$1 $k$-point mesh.\\

\section{Deformation potential}\label{Appendix:deformation_potential}
\begin{figure}[h]
    \centering
    \includegraphics[scale=0.8]{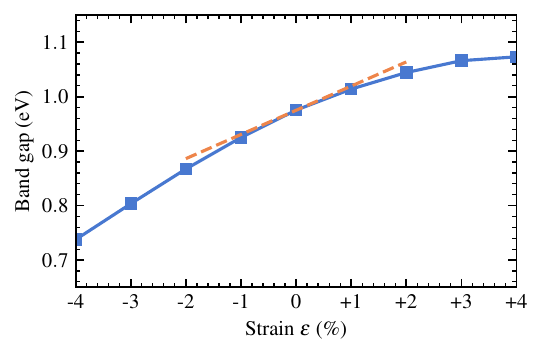}
    \caption{Bandgap under strain along the $a$ axis.}
\end{figure}
To estimate the in-plane coupling strength between excitons and acoustic phonons, we calculate the uniaxial deformation potential, $D=\Delta E_{\mathrm{g}}/\varepsilon$, where $\Delta E_{\mathrm{g}}$ is the change in the band gap due to a small strain. We simulate this strain by modifying the lattice parameter $a$ to be $1\%$ larger and smaller and relaxing the atomic positions. The band gap increases by 50.3~meV and decreases by $38.6$~meV with $1\%$ tensile and compressive strain which gives an average deformation potential of $D = 4.44$~eV. This calculation neglects excitonic effects that might modify the deformation potential $D$.

\end{document}